# Unsupervised COVID-19 Lesion Segmentation in CT Using Cycle Consistent Generative Adversarial Network

Authors: Chengyijue Fang[1#], Yingao Liu[1#], Jie Wen[2*], Yidong Yang [3,4*]

[#] the authors are contributing the same

[1] Department of Engineering and Applied Physics, University of Science and Technology of China, Hefei, Anhui, 230026 China

[2] Department of Radiology, The First Affiliated Hospital of USTC (Anhui Provincial Hospital). Division of Life Sciences and Medicine, University of Science and Technology of China, Hefei, Anhui, China

[3] Department of Radiation Oncology, the First Affiliated Hospital of USTC. Division of Life Sciences and Medicine, University of Science and Technology of China, Hefei, Anhui, 230026 China

[4] School of Physical Sciences & Hefei National Laboratory for Physical Sciences at the Microscale, University of Science and Technology of China, Hefei, Anhui, 230026 China

*Corresponding author:

Jie Wen, PhD

Professor

Department of Radiation Oncology, the First Affiliated Hospital of USTC

E-mail: jiewen@ustc.edu.cn

Yidong Yang, PhD, DABR

Professor and Director of Medical Physics Program

School of Physical Sciences, University of Science and Technology of China

E-mail: ydyang@ustc.edu.cn

## Abstract:

**Purpose:**

COVID-19 has become a global pandemic and is still posing a severe health risk to the public. Accurate and efficient segmentation of pneumonia lesions in CT scans is vital for treatment decision-making. We proposed a novel unsupervised approach using cycle consistent generative adversarial network (cycle-GAN) which automates and accelerates the process of lesion delineation.

**Method:**

The workflow includes lung volume segmentaion, "synthetic" healthy lung generation, infected and healthy image subtraction, and binary lesion mask creation. The lung volume volume was firstly delineated using a pre-trained U-net and worked as the input for the later network. The cycle-GAN was developed to generate synthetic "healthy" lung CT images from infected lung images. After that, the pneumonia lesions are extracted by subtracting the synthetic "healthy" lung CT images from the "infected" lung CT images. A median filter and K-means clustering were then applied to contour the lesions. The auto segmentation approach was validated on two public datasets (Coronacases and Radiopedia).

**Results:**

The Dice coefficients reached 0.748 and 0.730, respectively, for the Coronacases and Radiopedia datasets. Meanwhile, the precision and sensitivity for lesion detection are 0.813 and 0.735 for the Coronacases dataset, and 0.773 and 0.726 for the Radiopedia dataset. The performance is comparable to existing supervised segmentation networks and outperforms previous unsupervised ones.

**Conclusion:**

The proposed unsupervised segmentation method achieved high accuracy and efficiency in automatic COVID-19 lesion delineation. The segmentation result can serve as a baseline for further manual modification and a quality assurance tool for lesion diagnosis. Furthermore, due to its unsupervised nature, the result is not influenced by physicians' experience which otherwise is crucial for supervised methods.

# 1. INTRODUCTION

The coronavirus disease 2019(COVID-19) has become a global public health problem and now still affects billions of people's life. According to the World Health Organization, the pandemic has caused over 2 million deaths by April 2021[1]. The typical symptom of COVID-19 includes cough, fever, and pneumonia after infection[2]. Clinically, CT scans are commonly used to evaluate the progress and severity of pneumonia[3,4] due to their high resolution in three dimensions and broad availability compared to other imaging modalities. Accurate delineation of pneumonia lesions is vital for evaluating disease progression and assessing the severity of infection which is crucial for treatment decision making[5]. However, manual segmentation is time-consuming and labor-intensive. Therefore, automatic segmentation methods are highly demanded.

In the past decade, deep learning has shown its tremendous power and potential in various radiological applications, including image segmentation[6,7], disease classification and synthetic image generation[8–11]. Since the beginning of the COVID-19 pandemic, deep learning, incorporating various modalities of imaging techniques like CT, X-ray and Ultrasound[4,12–15], has been applied in clinical diagnosis, predicting disease progress[16], classifying pneumonia types and assessing severity of infection[4,17]. However, existing methods are mostly based on supervised learning which requires substantial data labeling by radiologists as training references. For example, U-net networks have been used for the classification and segmentation of COVID-19 lesions in CT scans[18,19,20]. The results vary greatly among different studies, partially due to the inter- and intra-observer variations in the training lesion labeling by different radiologists[18].

Compared to supervised learning, unsupervised learning does not require training labeling, and hence gets rid of the burden of manual lesion delineation and the inter- and intra-observer inconsistency. For example, Yao et al [21] proposed a label-free pneumonia lesion segmentation method which employed an unsupervised statistical method to simulate infected lungs from healthy ones. Zhang et al[22] developed an unsupervised method to augment lung images for better followup segmentation training using conditional GAN [22]. However, most of these methods focused on data augmentation with unsupervised network, and had to rely on supervised networks to train the lesion segmentation process[6,21].

Cycle consistent GAN (cycle-GAN) is an unsupervised network that has been widely used in medical image analysis, such as synthetic CT generation [23–25] and image transformation between different MRI sequences[9,24]. Inspired by this, we propose a cycle-GAN-based unsupervised framework for COVID-19 lesion segmentation. The cycle-GAN is used to convert infected lung slices to healthy lung slices by transforming pneumonia lesions into normal lung tissues. Then, the lesion is retrieved by subtracting the simulated "healthy" lung from the original image. The network does not require any image pairing or manual training label, hence can improve efficiency and eliminate the inter- and intra-observer inconsistency otherwise presented in supervised networks.

## 2. MATERIALS AND METHODS

### 2.A. Patient dataset

In this study, CT scans of 77 COVID-19 patients with positive reverse transcription polymerase chain reaction results are collected between Dec 2019 and Jan 2020 in the First Affiliated Hospital of University of Science and Technology of China. The data is anonymized before any analysis. The patient and CT scan information is listed in Table 1. All CT images are converted to 1x1x1 mm³ spatial resolution and cropped to 256x256 pixels per slice. The image window level is set to [-800 100] HU and all images are normalized to [-1, 1] with a zero background before being used for network training. We select 1264 healthy CT slices and 1272 slices with pneumonia lesions from 77 CT scans as training dataset. Both lungs are extracted from CT images firstly using a U-net segmentation network and the results are verified before further processing. The extracted lung images are used as the input of cycle-GAN.

Table 1: Patient and original CT scan information.

| Patient demographics | | CT scans | | |
| --- | --- | --- | --- | --- |
| Total number | 77 | Manufacturer | NEUSOFT | GE |
| Gender(male/female) | 45/32 | Number of scans | 75 | 2 |
| Age(mean±SD) | 45±16 | Thickness(mm) | 0.8-2 | 0.625,1.25 |

**2.B. Cycle-GAN Structure**

In training stage, we firstly use a cycle-GAN strategy to generate synthetic healthy lung CT slices. The network architecture is illustrated in Figure 1. We denote the infected lung CT slices by domain $X$ and healthy lung CT slices by domain $Y$, the probability distribution for each domain is referred as $P_x$ and $P_y$, respectively. The generator $G_{XY}$ denotes the mapping from domain $X$ to domain $Y$, and $G_{YX}$ denotes the mapping from domain $Y$ to domain X. $\tilde{X}, \tilde{Y}$ are synthetic "healthy" and "infected" lung slices. Two adversarial discriminators $D_X$ and $D_Y$ are used to distinguish real input images and synthesized images. The architecture of the generator is a U-Net variant and consists of 8 stages, as shown in Figure 2. Unlike the original U-net[7], instance normalization, which can better preserve the image details in image generation process, is applied immediately after each convolutional layer except for the last one. All convolution filters in the generator have a size of 3×3 pixels. We set the channel number of the first block as 64. In the encoder part, the width and height of the feature map are halved using convolution with a stride of 2 instead of max pooling. In the first four stages, the channel number is doubled after the feature map passes each layer, while in the last four stages the channel number is fixed to 512. All the feature maps in the encoder part are concatenated with their counterparts in the decoder part. The encoder and decoder parts are symmetric.

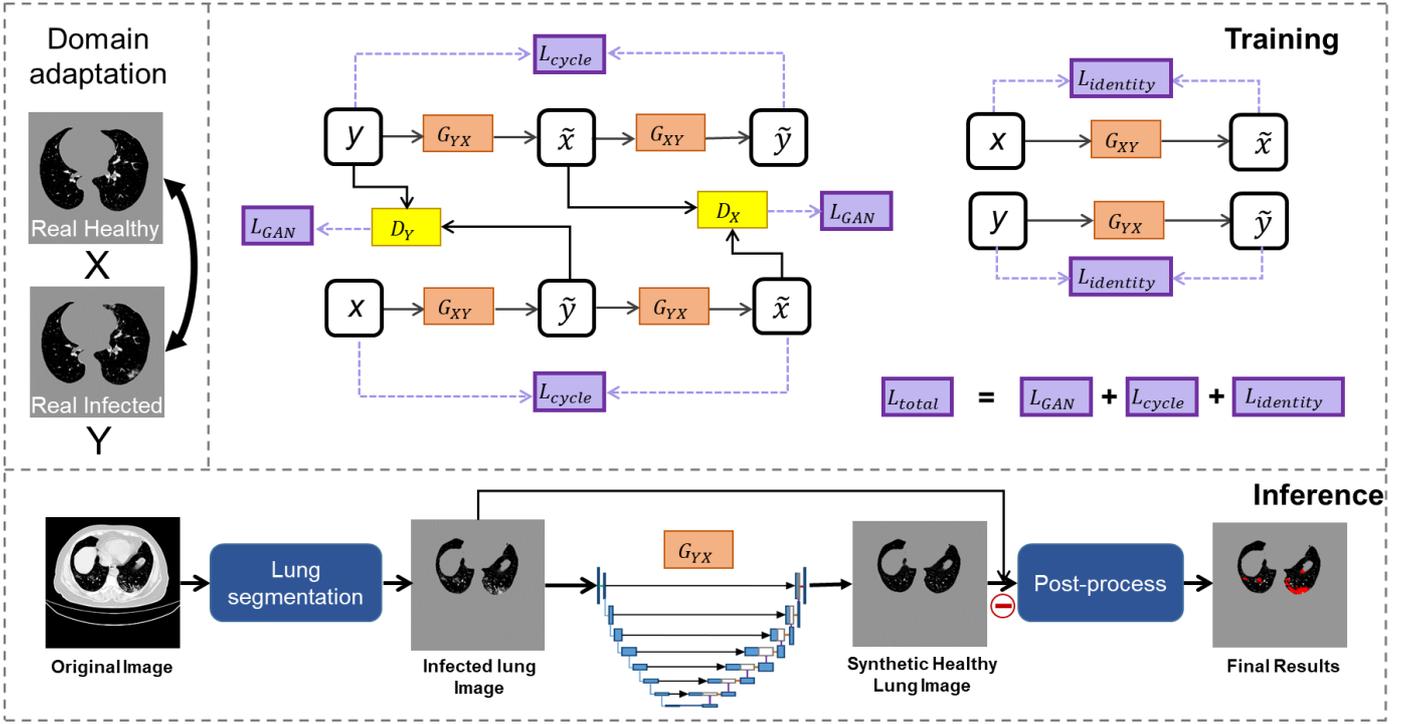

Figure 1: Scheme of the proposed method. $G_{xy}, G_{yx}$ are generators for generating synthetic "Healthy" images and "Infected" images. $l_{GAN}, l_{cycle}, l_{idtentity}$ are three loss functions used for training. $D_x$, $D_y$ are two discriminators that distinguish real and synthetic images.

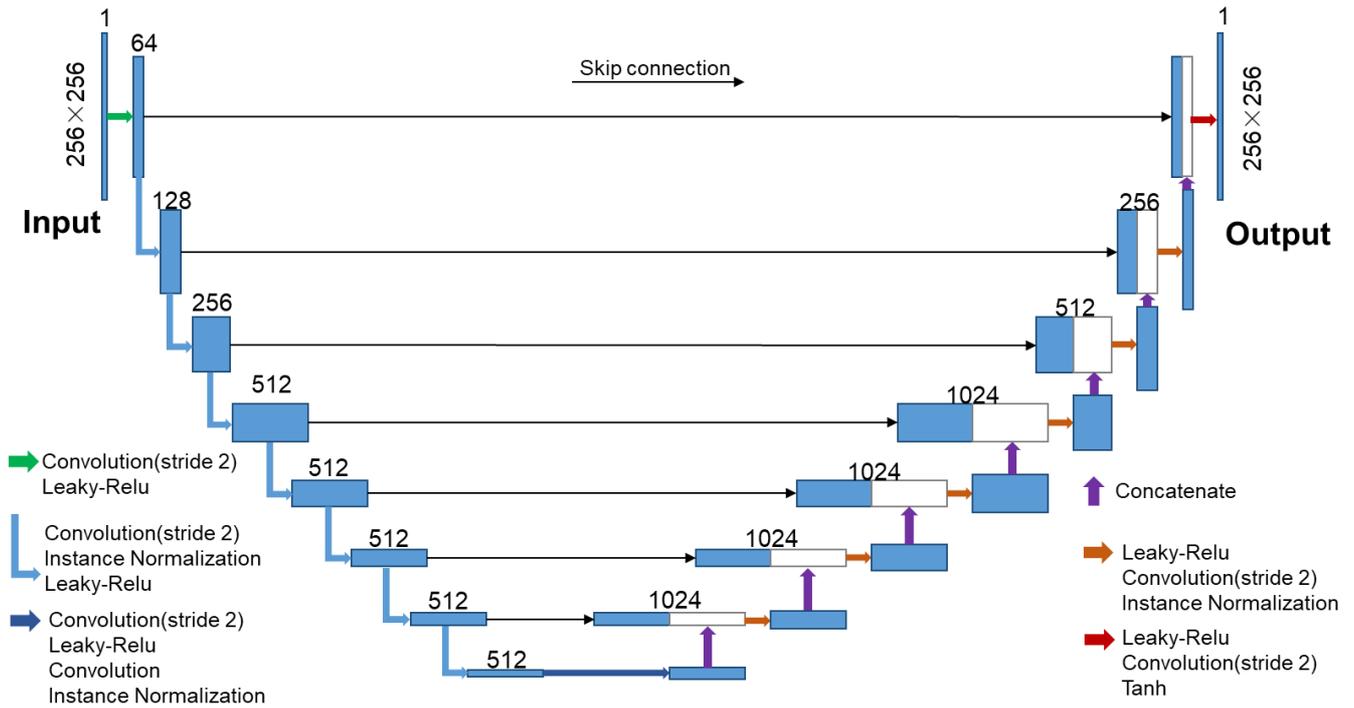

Figure 2: The structure of U-net Generator. The size of input and output image is 256×256. Encoder and decoder part each has 8 stages. Skip connection is applied to each stage.

The discriminator $D_X$ and $D_Y$ are implemented by a 70 × 70 Patch-GAN[26]. The architecture of the discriminator is illustrated in Figure 3. The first three convolution layers use a stride of 2, while the remaining convolution layers use a stride of 1. All the convolution layers have a padding of 1 and employ the Leaky-ReLU activation function with a slope parameter of 0.2. In the first convolution layer, a feature map with 64 channels is generated. After that, the channel number is doubled after the feature map passes each layer. In the last layer, the output is reduced to one channel.

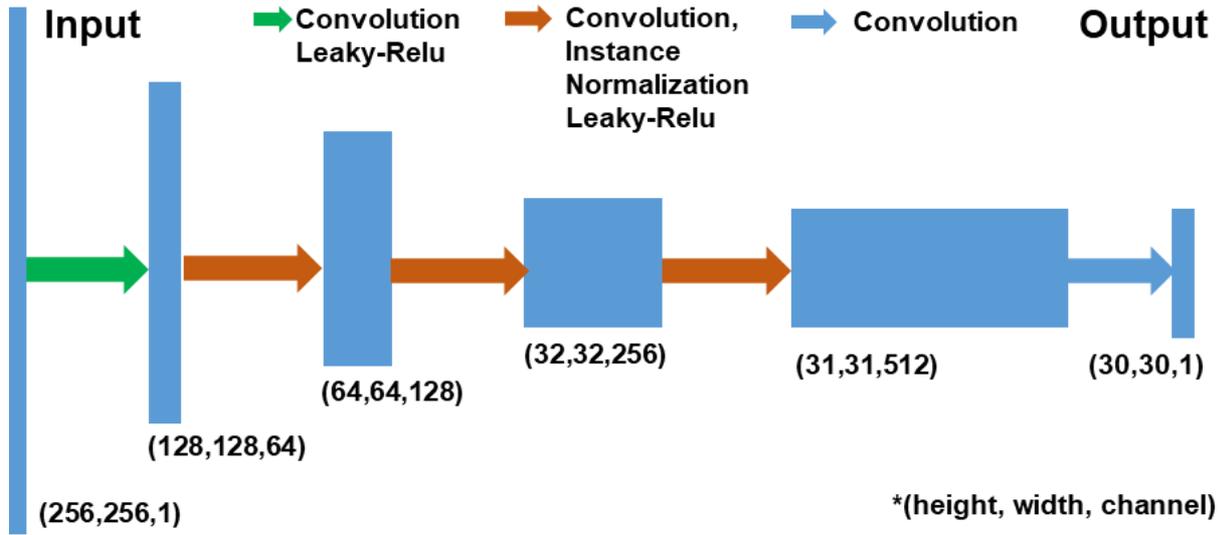

Figure 3: The Structure of Discriminator. In the first layer, convolution and leaky-Relu are applied. In the 2nd -4th layer, convolution, instance normalization and Leaky-Relu are implemented. Only one convolution layer is applied in the last layer.

The generators and discriminators are trained by solving a min-max problem : $\min_{G_{XY},G_{YX}} \max_{D_X,D_Y} l_{total}(G_{XY}, G_{YX}, D_X, D_Y)$ where $l_{total}(G_{XY}, G_{YX}, D_X, D_Y)$ is defined as:

$$l_{total}(G_{XY}, G_{YX}, D_X, D_Y) = l_{GAN}(G_{XY}, D_Y, X, Y) + l_{GAN}(G_{YX}, D_X, Y, X) +$$

$$\lambda_{cycle} l_{cycle}(G_{XY}, G_{YX}) + \lambda_{identity} l_{identity}(G_{XY}, G_{YX}) \qquad (1)$$

where $l_{total}$ the total loss aiming to learn the mapping function between the source and target domain. $\lambda_{cycle}$ and $\lambda_{identity}$ are introduced in (1) mainly to weigh the importance of the three losses. After optimization, we set $\lambda_{cyc} = 10$ and $\lambda_{id} = 5$.

$l_{GAN}$ is the loss function of the discriminator calculating the difference between synthetic "healthy" slices and real healthy slices. To maintain stability during the learning process, we here choose L2 loss in the LSGAN[27] as our loss function instead of the sigmoid cross entropy in regular GANs[28]. The $l_{GAN}$ is defined as :

$$l_{GAN} = E_{x \sim P_x}[(D_Y(G_{YX}(x)) - 1)^2] + E_{y \sim P_y}[D_Y(y)^2] + E_{y \sim P_y}[(D_X(G_{XY}(y)) - 1)^2]$$

$$+ E_{x \sim P_x}[D_X(x)^2] \qquad (2)$$

where $x \sim P_x$ denotes the learning process on domain X. $l_{cycle}$ is used to keep the consistency of the two generators $G_{XY}$ and $G_{YX}$, and is defined as :

$$l_{cycle}(G_{XY}, G_{YX}) = E_{x \sim Px}[\|G_{YX}(G_{XY}(y)) - x\|_1 + E_{y \sim Py}[\|G_{XY}(G_{YX}(x)) - y\|_1]$$

(3)

where $\|\cdot\|$ is the $l_1$ norm. Since we only want to convert unhealthy lung CT slices into healthy ones, an identity loss $l_{identity}$ in (1) is designed to keep the image feature when a healthy slice is input into the generator. The identity loss is defined as :

$$l_{identity}(G_{XY}, G_{YX}) = E_{y \sim Py}[\|G_{XY}(y) - y\|_1] + E_{x \sim Px}[\|G_{YX}(x) - x\|_1] \qquad (4)$$

We use the ADAM optimization method to train all the networks with $\beta_1 = 0.5$ and $\beta_2 = 0.999$. Kernels are initialized randomly with a Gaussian distribution. We update the generator and the discriminator at each iteration. The image slice is randomly cropped to patches of 256 × 256 pixel size as the input. The number of mini-batches is one, and the number of epochs is 100. The learning rate is initially set to 0.0002 and linearly decreased to 0 in the last 50 epochs. We stopped the training at the 85[th] epoch which had smallest loss for optimal performance.

**2.C. Image Post-processing**

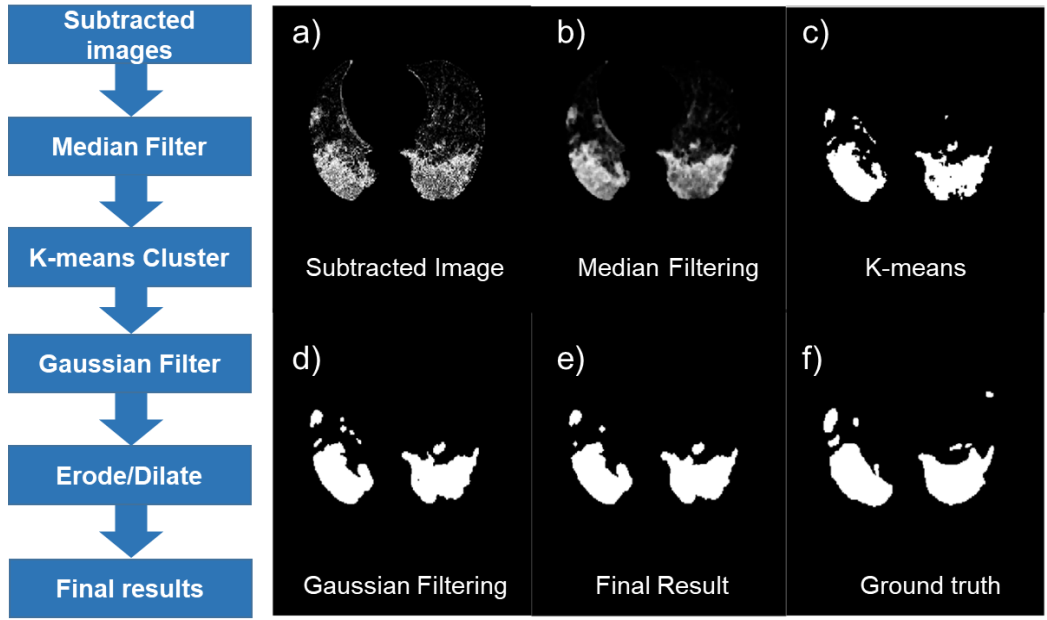

Figure 4: The framework of post-processing. (a)-(e) shows the image after subtraction of the synthetic "healthy" image from infected image (a), median filtering (b), K-means clustering with zero background (c), gaussian filtering (d), and hole filling, erosion and dilation. (f) is the ground truth.

The post-processing steps are illustrated in Figure 4. The synthetic healthy image is subtracted from its corresponding real infected image to obtain a difference map. A median filter is applied to the difference map to suppress noise and remove small islands. Then k-means clustering is used to segment the lesion from the low-intensity background. Finally, a 5x5 Gaussian kernel is employed to smoothen the lesion edge before erosion and dilation with a radius of 1 pixels are implemented to further remove small and isolated regions. The post-processing procedures are done in MATLAB 2018A and take less than 2 minutes for each patient. The training is conducted on a Linux machine with an NVIDIA RTX 2080Ti GPU and took 8 hours. As a comparison, we also compare the k-means clustering method with Otsu thresholding, which is commonly used for thresholding segmentation.

## 2.D. Segmentation Evaluation

In this study, we choose Coronacases and Radiopedia which are two public COVID-19 CT image databases to evaluate the performance of our method, with each database providing 10 and 9 patients with lesions delineated. The

Dice similarity coefficient (DSC), volume precision (PSC), and volume sensitivity(SEN) are used to evaluate the performance of the proposed segmentation method, which are defined as:

$$DSC(\%) = \frac{2(V_{pre} \cap V_{gt})}{V_{pre} + V_{gt}} \times 100\% \qquad (5)$$

$$PSC(\%) = \frac{V_{pre} \cap V_{gt}}{V_{pre}} \times 100\% \qquad (6)$$

$$SEN(\%) = \frac{V_{pre} \cap V_{gt}}{V_{gt}} \times 100\% \qquad (7)$$

where $V_{pre}$, and $V_{gt}$ represent the predicted and ground truth lesion volume.

Table 2: The comparison between the proposed and existing methods.

| Methods | Coronacases Database | | | Radiopedia Database | | |
|---|---|---|---|---|---|---|
| | DSC(%) | PSC(%) | SEN(%) | DSC(%) | PSC(%) | SEN(%) |
| nnUNet-2D | 86.9±3.3 | 86.7±5.5 | **87.3±2.9** | 87.1±5.4 | 92.6±3.6 | 82.4+7.8 |
| nnUNet-3D | **86.1±5.1** | 88.8±8.8 | 84.2±5.6 | **85.8±4.6** | **91.0±3.9** | **81.7±8.9** |
| COPLE-Net | 76.7±8.2 | 79.5±14.9 | 75.9±7.6 | 77.8±6.7 | 83.1±14.9 | 75.6±10.1 |
| Inf-Net | 59.5±16.8 | 59.4±19.2 | 65.2±22.2 | 64.8±12.0 | 72.6±17.2 | 63.0±18.2 |
| Label-free | 68.7±15.8 | **85.1±7.0** | 62.1±22.8 | 59.4±17.4 | 60.4±19.7 | 61.8±18.4 |
| Proposed (Otsu) | 70.9±14.3 | 68.2±21.1 | **81.1±9.1** | 65.3±12.6 | 65.5±15.6 | 70.6±16.6 |
| Proposed(k-means) | **74.8±12.1** | 81.3±8.79 | 73.5±20.5 | **73.0±9.5** | 77.3±17.8 | **72.6±11.1** |

## 3. RESULTS

As shown in Table 2, we compare our method with several existing supervised and unsupervised methods. We trained nn-Unet[29] both in 2D and 3D fomats using all the data of Coronacases and Radiopedia in five folds. Meanwhile, we directly used another supervised network COPLE-Net[18] and a semi-supervised network Inf-Net[30] using different datasets. The results of our method reach a dice coefficient of 74.8±12.1 and 73.0± 9.5 on Coronacases and Radiopedia, respectively. The precision and sensitivity is 81.3±8.79, 73.5±20.5 in Coronacases and 77.3±17.8, 72.6± 11.1 in Radiopedia. As shown in table 2, our approach is comparable with supervised COPLE-Net method and outperforms the semi-supervised Inf-Net method. We also compare our method with a state-of-the-art unsupervised label-free[21] method (denoted as "label-free" in Table 2). Our method results in higher scores in most indices and is more robust than the "label-free" method when implemented between different datasets. The results also indicates that k-means clustering performs better than Ostu thresholding in image post-pocessing.

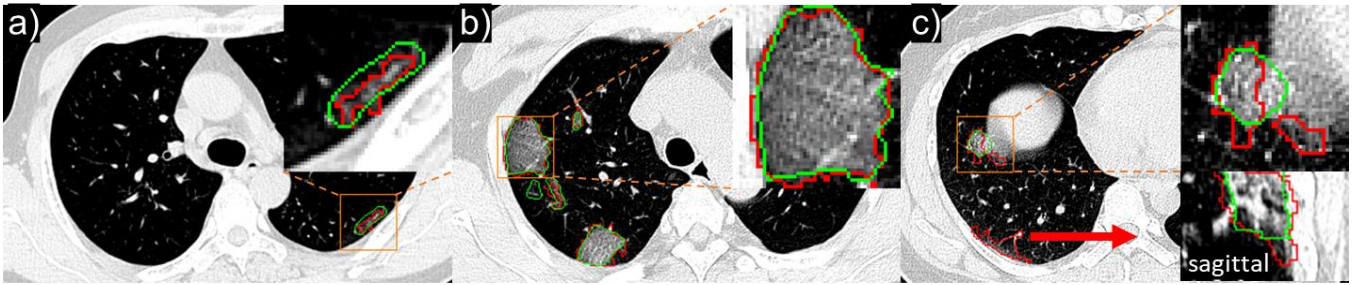

Figure 5: Segmentation of various size lesions. (a)-(c) are segmentation images from different patients. The upper right panel is the zoom-in of the boxed area in each image. The red outlines are from the proposed method, and the green are ground truth. The red arrow in (c) points to the sagittal plane across the lesion.

The proposed method performs well in small lesion segmentation. Figure 5(a) shows that a small lesion with only 2mm width is correctly delineated. As shown in figure 5(b), our method can also readily separate lesions from the chest wall. Interestingly, our method catched some low contrast lesions which are skipped by radiologists during manual segmentation, as shown in figure 5(c).

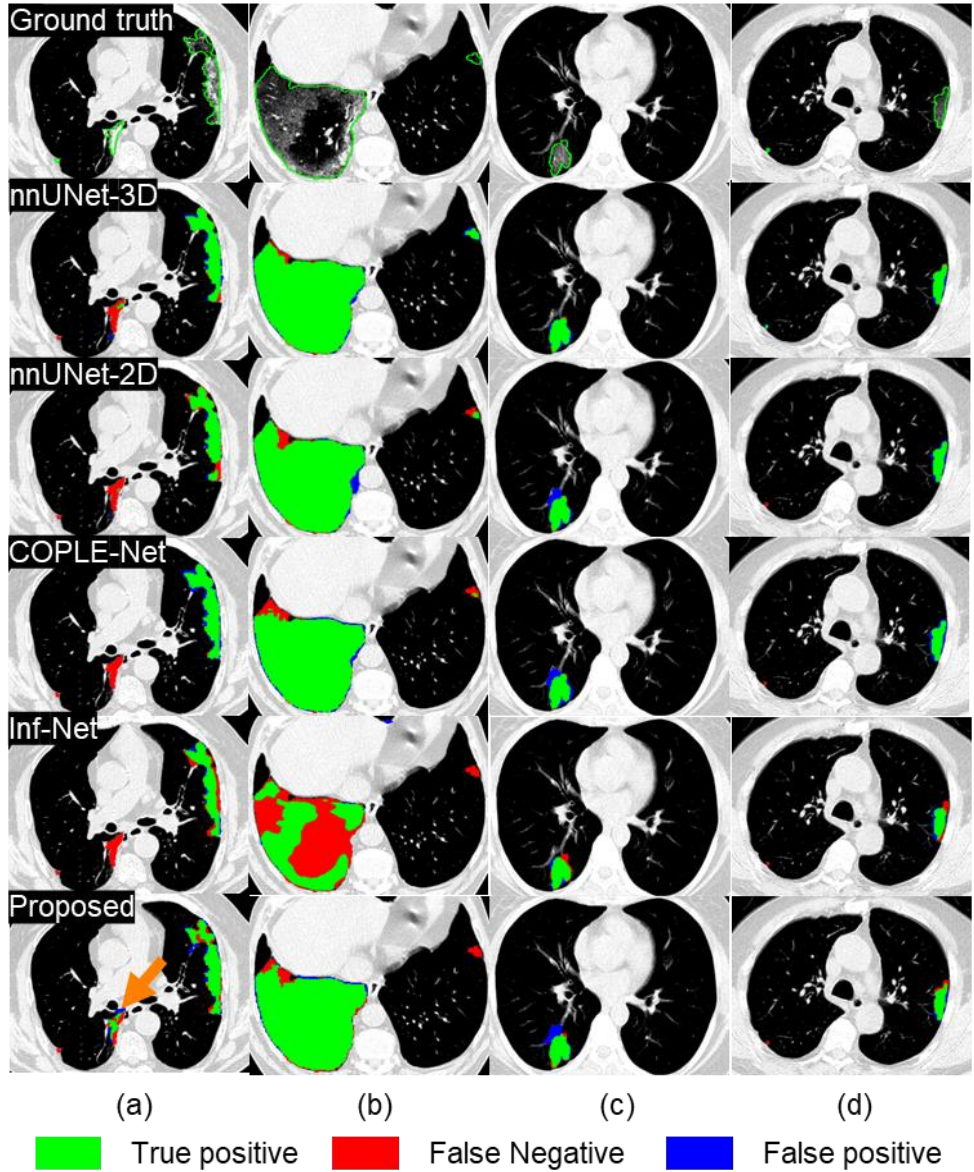

Figure 6: Visual comparsion of different lesion segmentation methods. (a)–(d) show segmentation results from four different patients. The green, red, and blue areas are true positive, false negative and false positive, respectively.

Figure 6 demonstrates the performance of our method compareing to existing state-of-art supervised methods. These lesions are various in shape, size, and position. As shown in figure 6(a), our method localized and delineated, while other methods missed, the lesion close by the trachea. Our method performs better than Inf-Net and COPLE-Net on a large lesion segmentation as shown in figure 6(b), while is comparable to all existing supervised methods in segmenting small lesions isolated in the lung volume and nearby the chest wall as shown in figure 6(c) and 6(d).

To further evaluate the capability of our method on lesion diagnosis, we divided the lung volume into 12 subvolume as illustrated in figure 7 and counted whether there are lesions in each region. We compared our method with supervised methods using evaluation metrics including diagnostic accuracy, precision and sensitivity. As shown in Table 3,the accuracy of our method reached 93% on Coronacases and 86% on Radiopedia dataset, which is comparable with the supervised methods.

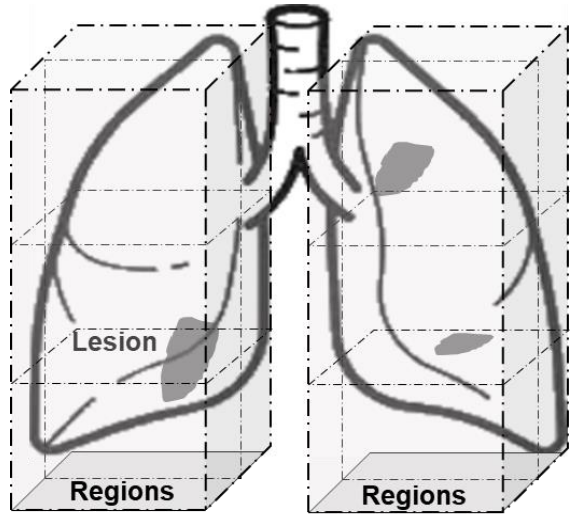

Figure 7: The lung volume is divided 12 regions equally.

Table 3: Lesion region diagnosis accuracy, precision and sensitivity by dividing patient into 12 regions

|  | Coronacases | | | | |
| --- | --- | --- | --- | --- | --- |
|  | nnUnet-3D | nnUnet-2D | Cople-Net | Inf-Net | Proposed |
| **ACC%** | **0.93** | 0.81 | 0.83 | 0.70 | **0.93** |
| **PSC%** | **0.73** | 0.39 | 0.46 | 0.05 | 0.72 |
| **SEN%** | 1.00 | 0.99 | 0.99 | 1.00 | 0.98 |
|  | Radiopedia | | | | |
| **ACC%** | 0.91 | 0.71 | 0.93 | **0.93** | 0.86 |
| **PSC%** | 0.78 | 0.65 | 0.10 | 0.78 | 0.69 |
| **SEN%** | 0.97 | **1.00** | 0.97 | 0.97 | 0.89 |

## 4. DISCUSSION

In this study, we propose an unsupervised method for delineating COVID-19 lesions using cycle-GAN. This unsupervised learning method shows great potential in lesion segmentation without employing labeled data, and is validated on different public database. It can work as an efficient and independent automatic segmentation method or provide a start point for physicians followup refinement.

Table 4: Results of nn-UNet when trained on Coronacases dataset and test on Radiopedia dataset.

| Methods | DSC(%) | PSC(%) | SEN(%) |
| --- | --- | --- | --- |
| nnUNet-2D | **70.1±10.0** | 67.7±21.0 | 79.4±10.6 |
| nnUNet-3D | 69.5±14.0 | **74.1±21.9** | 72.4±12.4 |
| Proposed | 68.7±13.0 | 62.4±15.7 | **80.1±16.6** |

The proposed method is robust and less database dependent. Different datasets are labeled by different radiologists, and there may exist considerable labeling inconsistency heavily depending on physicians' experiences and habits. We trained nnUnet-2D and -3D on Coronacases database using 5-fold cross-validation, and then tested on Radiopedia database. The results, as shown in Table 4, are much worse than that when the network was trained on mixed database in table 3, attributing to the inconsistency between the training and testing datasets. We also trained our proposed method on Coronacases and tested on Radiopedia. As shown in table 4, the results of our proposed method is less influenced by the inconsistency between different dataset.

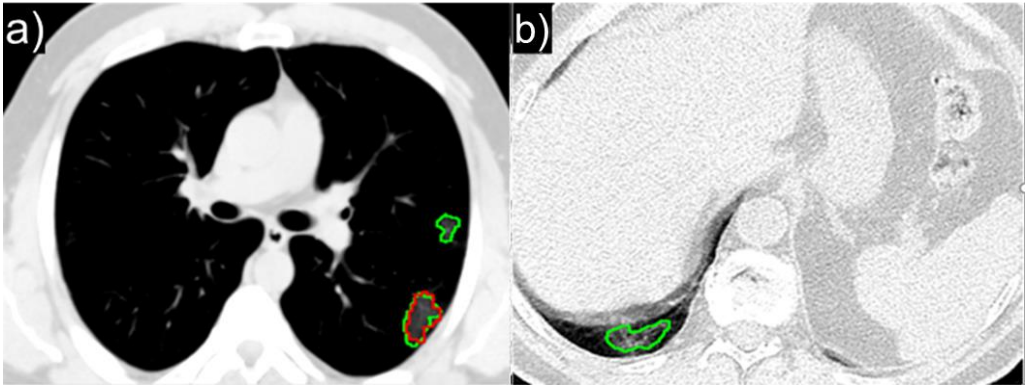

Figure 8: Limitations of our method. The green contours are mannaul labels and the red ones are results of our method.

There are some limitations for the proposed unsupervised method. The method missed some small and low contrast lesions, as shown in Figure 8(a). The sample number might not be sufficiently large, particularly lacking patients with lesions in the peak superior and bottom inferior parts of the lung volume, resulting in missed delineation as shown in Figure 8(b). In this work, we trained the network with 2D images due to hardware limitation, which didn't make full use of the three-dimension image property. Future study using 3D image input may further improve segmentation accuracy, particularly the contour continuity along the image thickness direction.

For the post-processing, we only used Otsu thresholding and k-means methods which are two simple but common methods. In spite of this, the proposed unsupervised method still achieved decent segmentation results, with a dice value of 0.748 and accuracy of 0.93 in Coronacases database. In the future, the unsupervised method can be combined with more sophysicated post-processing methods, such as texture analysis or even additional deep learning network, to further improving segmentation results.

## 5. CONCLUSIONS

In this work, we propose an unsupervised approach that can accurately and efficiently delineate the COVID-9 lesions automatically in CT scans. The traing process of the unsupervised network does not rely on any labeled data. The segmentation result can serve as a baseline for further manual modification and a quality assurance tool for lesion diagnosis. Furthermore, due to its unsupervised nature, the result is not influenced by physicians' experience which otherwise is crucial for supervised methods.

## ACKNOWLEDGEMENTS

Research reported in this publication is supported by the Fundamental Research Funds for the Central Universities (Grant No. WK2030000037), Anhui Provincial-level S&T Megaprojects (Grant No. BJ2030480006), and partially supported by "USTC Research Funds of the Double First-Class Initiative" (YD9110002011), the National Natural Science Foundation of China (22077116) and the Collaborative Innovation Program of Hefei Science Center, CAS (2020HSC-CIP010).